\documentclass[fleqn,usenatbib]{mnras}

\usepackage[utf8]{inputenc}

\usepackage{acronym}
\usepackage{ae,aecompl}
\usepackage{booktabs}
\usepackage{dcolumn}
\usepackage{graphicx}
\usepackage{amsmath,amsfonts,amssymb}
\usepackage{mathrsfs}
\usepackage[normalem]{ulem}
\usepackage{epstopdf}
\usepackage[T1]{fontenc}

\let\oldhref\href
\renewcommand{\href}[2]{\oldhref{#1}{\hbox{#2}}}

\usepackage{color}
\definecolor{cyan}{rgb}{0, 0.7, 1}

\title[Sensitivity of presupernova neutrinos to stellar evolution models]{The sensitivity of presupernova neutrinos to stellar evolution models}

\author[C. Kato et al.]{
Chinami Kato$^{1}$\thanks{E-mail:chinami.kato.e8@tohoku.ac.jp},
Ryosuke Hirai$^{2,3,4}$,
Hiroki Nagakura$^{5}$
\\
$^{1}$Department of Aerospace Engineering, Tohoku University, 6-6-01 Aramaki-Aza-Aoba, Aoba-ku, Sendai 980-8579, Japan\\
$^{2}$OzGrav: Australian Research Council Centre of Excellence for Gravitational Wave Discovery, Clayton, VIC 3800, Australia\\
$^{3}$Monash Centre for Astrophysics, School of Physics and Astronomy, Monash University, Clayton, VIC 3800, Australia\\
$^{4}$Department of Physics, University of Oxford, Keble Rd, Oxford, OX1 3RH, United Kingdom\\
$^{5}$Department of Astrophysical Sciences, Princeton University, 4 Ivy Lane, Princeton, NJ 08544, USA
}

\date{Accepted XXX. Received YYY; in original form ZZZ}

\pubyear{2020}

\begin{document}
\label{firstpage}
\pagerange{\pageref{firstpage}--\pageref{lastpage}}
\maketitle

\begin{abstract}
We examine the sensitivity of neutrino emissions to stellar evolution models for a 15$M_\odot$ progenitor, paying particular attention to a phase prior to the collapse.
We demonstrate that the number luminosities in both electron-type neutrinos ($\nu_e$) and their anti-partners ($\bar{\nu}_e$) differ by more than an order of magnitude by changing spatial resolutions and nuclear network sizes on stellar evolution models. 
We also develop a phenomenological model to capture the essential trend of the diversity, in which neutrino luminosities are expressed as a function of central density, temperature and electron fraction. In the analysis, we show that neutrino luminosity can be well characterized by these central quantities. This analysis also reveals that the most influential quantity to the time evolution of $\nu_e$ luminosity is matter density, while it is temperature for $\bar{\nu}_e$.
These qualitative trends will be useful and applicable to constrain the physical state of progenitors at the final stages of stellar evolution from future neutrino observations, although more detailed systematic studies including various mass progenitors are required to assess the applicability.
\end{abstract}


\section{Introduction}\label{sec:intro} 
Growth of detection capabilities for low-energy ($\lesssim$ a few tens of MeV) neutrinos has played a prominent role in our understanding of stellar astrophysics. The theory of solar evolution 
may be established if neutrino oscillation models have provided a solution to the solar neutrino problem, in which the deficit in observed electron-type neutrino ($\nu_e$) from theoretical predictions can be complemented by neutrino flavor conversions \citep{ahmad2001,ahmad2002}. Another big breakthrough in neutrino-stellar astrophysics came with the direct detection of neutrinos associated with a close proximity core-collapse supernova (CCSN), SN1987A \citep{hirata1987,Bionta1987,Alekseev:1987ej}, which has provided precious information on the CCSN dynamics during the development of the explosion.


Neutrino observations associated with stellar dynamics are expected to evolve considerably over the next few decades with refined/improved detection techniques, especially in the few MeV energy regime. In the SK-Gd project, currently in progress at Super-Kamiokande, gadolinium is doped into pure water which leads to a substantial reduction of background noise while sustaining the capability of the detector response \citep{Beacom2004,Simpson2019}. Liquid scintillator detectors with improved sensitivities for the few MeV neutrinos are also being developed in Borexino \citep{borexino2009}, KamLAND \citep{gando2013,gando2015}, SNO+ \citep{sno+2015} and JUNO \citep{an2016,han2016}. Other detectors targeting not only electron-type anti-neutrinos ($\bar{\nu}_e$) but also other species are under way; for instance, DUNE \citep{acciarri16} detecting $\nu_e$ through charged-current reactions with liquid argon, or large-scale dark matter experiments such as Darkside-20k \citep[liquid argon,][]{aalseth2018}, DARWIN \citep[liquid xenon,][]{aalber2016} and superCDMS \citep[germanium and silicon,][]{agnese2016}, detecting all types of neutrinos through coherent elastic neutrino-nucleus scattering.


Enhancing detector sensitivities extends the threshold distance up to which sources can be detected. These improvements are particularly important for ``presupernova neutrinos'', which are neutrinos emitted at the last stage of stellar evolution up to the onset of gravitational collapse. Roughly speaking, the total number luminosity of neutrinos reaches $\sim 10^{55} {\rm s^{-1}}$ with a few MeV average energy at the onset of core collapse \citep{odrzywolek2004,Kato2015,Yoshida2016, Kato2017,Patton2017b,Patton2017}, implying that it is detectable in the currently available detectors only if sources are located at $\lesssim$ 1 kpc, i.e., the corresponding event rate is quite low \citep{Asakura2016,Simpson2019,Li2020}. 
It should be noted that the sensitivity of detectors will become better and hence it is meaningful to consider how we can maximally take advantage of up-coming presupernova neutrino events to study the end-stages of the lives of stars.

Observations of presupernova neutrinos will play three major roles: supernova warning alarm (SN alarm), constraining neutrino physics and providing insight to the theory of stellar evolution. The time scale of presupernova neutrino emission is governed by that of the stellar evolution, indicating that the neutrino signals will reach us more than days before CCSN explosion. It will be, hence, the first signal coming from the progenitor. In the wake of the SN alarm, other observational instruments such as for gravitational waves and electromagnetic waves  can be prepared for the upcoming burst signal \citep{Kato2017,Patton2017b,Raj2019,Asakura2016,Simpson2019,Li2020}. By virtue of the high sensitivity of presupernova neutrinos to the oscillation models, it may be possible to determine the neutrino mass hierarchy through $\nu_e$ events \citep{Kato2017} or by analyzing multiple reaction channels in detectors such as inverse-$\beta$ decay and electron scattering \citep{Guo2019}. Note that effects of collective neutrino oscillation, which are one of the major uncertainties to determine the flavor conversion in CCSN neutrinos, can be neglected for presupernova neutrinos, i.e., the oscillation model is very simple (but see \cite{lunardini2001} for Earth matter effects). Finally, presupernova neutrinos will provide direct information on the physical state deep inside of the progenitor, which is the main topic of this paper.







Although the overall theoretical picture of the late evolution stages of massive stars has already been established, the details of these final phases are extremely uncertain. The lack of direct observation of the progenitors of these phases is one of the major obstacles. They are
due to the short nuclear timescale towards the end of the lives of massive stars; for instance, the core-C burning stage only lasts for several centuries and then reaches core-collapse within several years after that. This is many orders of magnitude shorter than the average lifetime of massive stars, making it extremely difficult to find stars in this stage. In addition to this, the nuclear timescale is much shorter than the thermal timescale of the star after core-C burning, meaning that physical properties at the stellar surface may be insensitive to any changes in the core, i.e., it is very hard to catch any symptoms of very last stage of stellar evolution from electromagnetic observations. We also note that, after CCSN explosions, the central chemical composition at the final stages of massive stars can be probed through spectroscopic observations of CCSN ejecta \citep{tolstov2019,john2019}, which would be informative to constrain the progenitor properties. However, the chemical composition may be substantially affected by the explosive nucleosynthesis during the explosion, indicating that the information is limited.

It is even more challenging to understand the late phases of stellar evolution from the theoretical side. Current theoretical studies of stellar evolution are mainly based on 1D stellar models with special treatments to account for multi-dimensional effects such as convection or rotation. Recent multi-dimensional simulations of O-burning stars suggest that the 1D treatments (mixing length theory) actually describe the physics well \citep{mueller2016,yoshida2019}.  However, it is also known that the 1D stellar evolution codes are not consistent among themselves, with slightly different implementations of convection, overshooting, nuclear burning, mass-loss, etc. Even the slightest variations in these treatments can strongly affect the structure of the convective layers, which in turn affects the mixing of chemical elements. In fact, the problem is so subtle that the evolution can be substantially different just by slightly changing the spatial resolution \citep{farmer2016,sukhbold2018}.

We expect that presupernova neutrinos will be a direct and unique observable carrier to bring detailed information on the physical state deep inside the core at the last stage of stellar evolution, and also being a smoking gun to build its theory comprehensively. Indeed, neutrinos can propagate through stars freely and be detected without losing the 
information of the core. Hence, the neutrino signals can potentially be used to place constraints on progenitor types \citep{Kato2015}, shell-burning properties \citep{Yoshida2016} and $\beta$ process \citep{Patton2017}. It should be noted, however, that previous studies on presupernova neutrinos have been carried out under given stellar models, whereas the theoretical model is not definitive as mentioned above.
In fact, very little is known about how sensitive presupernova neutrinos reflect the change of input physics and numerical treatments on computations of stellar evolution.
These uncertainties will smear our understanding in the connection between the presupernova neutrino signals to the stellar evolution in real observations.

In this paper, we present results of the first systematic study for the sensitivity of presupernova neutrinos on stellar evolution models, focusing on a 15 $M_\odot$ CCSN progenitor. We generate 20 stellar evolution models with a stellar evolution code, MESA, by changing its nuclear network size and spatial resolutions and then evaluate the neutrino luminosity on each model by following the same procedure as in \citet{Kato2015, Kato2017}. We address some fundamental questions regarding a connection between presupernova neutrinos and stellar evolution models: how sensitive neutrino signals are to differences in stellar evolution models; what stellar quantity is the most influential to presupernova neutrino emission.
We analyze this sensitivity by developing a phenomenological approach, in which neutrino luminosity is expressed as a function of central density, temperature and electron fraction. Such a simple but useful approach will be applicable to interpret neutrino signals in future observations.

This paper is organized as follows: In Section~\ref{sec:model} we briefly summarize the stellar evolution models and describe methods for calculating neutrino signals. The diversity of presupernova neutrinos induced by different stellar evolution simulations is shown in Section~\ref{sec:variety}, and then we also quantify the impact of different treatments of computing nuclear compositions in Section \ref{sec:NSE}. Our phenomenological approach to capture the qualitative trend of the diversity of presupernova neutrinos is presented in Section \ref{sec:correlation}. Finally we conclude our study with summary and discussion in Section \ref{summary}.

\begin{figure}
    \centering
    \includegraphics[width=\columnwidth]{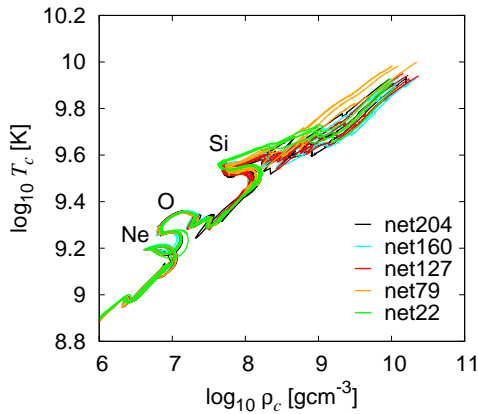}
    \caption{The evolutionary tracks of the central density and temperature for the 20 progenitor models. Colors distinguish network sizes for stellar evolution calculations.
    Some major nuclear burning stages are marked with labels.}
    \label{rhoc_tc}
\end{figure}

\begin{figure}
    \centering
    \includegraphics[width=\columnwidth]{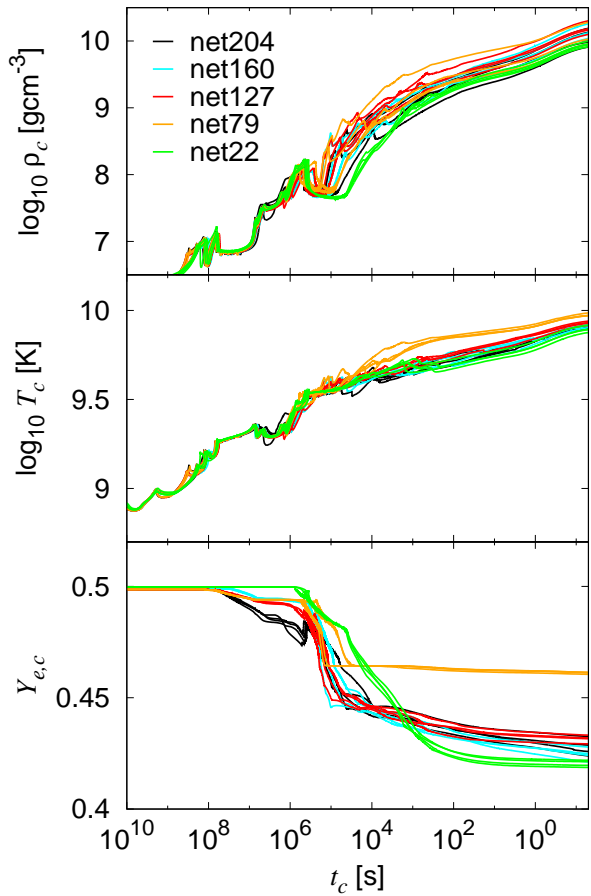}
    \caption{The time evolutions of the central density (top), temperature (middle) and electron fraction (bottom) for the 20 progenitor models.
    Colors distinguish network sizes for stellar evolution calculations.}
    \label{time_center}
\end{figure}

\section{Methods and models} \label{sec:model}
\subsection{Progenitor models}

We simulate the stellar evolution of 15~$M_\odot$ progenitors up to core collapse using the 1D stellar evolution code MESA \citep{paxton2011,paxton2013,paxton2015}, that corresponds to the same models as those in \cite{farmer2016}.
They investigate the dependence of spatial resolution, nuclear network and mass-loss on the evolution of 15, 20, 25 and 30$M_\odot$ progenitors.
Out of them, we select 20 models with 15 $M_\odot$ without mass-loss in this paper.
They have different numerical sets in the maximum grid size in mass coordinate\footnote{We note that this criterion does not influence the spatial resolution towards the centre of the star, where the resolution is determined by other criteria.} $\Delta M_{\rm{max}}$ = 0.1, 0.05, 0.02, 0.01$M_\odot$ and the number of nuclear species $N_{\rm{net}}$ = 22, 79, 127, 160, 204.
The calculations are terminated when the in-fall velocity exceeds 1000 km/s.
We ask readers to refer to \cite{farmer2016} about other technical settings in the MESA code. For all models we have utilised the input files provided in the MESA market place by the authors\footnote{http://cococubed.asu.edu/mesa\_market/}. We have noticed, however, that the $N_{\rm{net}}$ = 160 network in the repository (\texttt{mesa\_160.net}) was different from the one that was used in \citet{farmer2016} (\texttt{mesa\_160\_new.net}). For the purpose of the consistency among our models, we utilised the former network (\texttt{mesa\_160.net}) in our models.
We also note that the model with $(\Delta M_{\rm{max}},N_{\rm{net}})$ = (0.1,204) is the same progenitor as that employed in \cite{Patton2017b,Patton2017} and we employ this model to see the sensitivity of neutrino emissions to different treatments of computing nuclear compositions (see Section~\ref{sec:NSE}).

Figure~\ref{rhoc_tc} shows the evolutionary paths of the central density and temperature for the 20 progenitor models,
in which colors distinguish network sizes for stellar evolution simulations.
Some major nuclear-burnings are marked with labels.
In all the models, the core experiences H, He, C, Ne, O and Si-burnings and an Fe-core is formed, finally.
We, however, find that the central density and temperature are different from each other, even though the initial conditions are identical among models.
Their diversity is especially prominent after Si-burning ($\rho_c \sim 10^8\ \rm{gcm^{-3}}$), which can also be confirmed in the time evolutions of the central quantities displayed in Figure~\ref{time_center}.
They separate at $t_c < 5\times10^5$ s, which corresponds to the initiation of Si-burning.
Here $t_c$ is defined as time to the onset of core collapse. Note that the central quantities at the point of core collapse is also diverse; the central density (top), temperature (middle) and electron fraction (bottom) at $t_c$ = 0 have ranges of $10^{9.96} \leq \rho_c/\rm{gcm^{-3}} \leq 10^{10.37}$, $0.674 \leq T_c/\rm{MeV} \leq 0.861$ and $0.418 \leq Y_e \leq 0.461$, respectively. This represents a stochastic nature of late evolution stage of massive stars \citep{sukhbold2018} and has considerable influence on presupernova neutrino signals as demonstrated in the following sections.

\subsection{Methods for computing neutrino signals} \label{sec:neutrino}

Given matter profiles simulated by MESA, we compute neutrino number luminosities by a post-processing manner. In this study, six weak-processes are included:
\begin{eqnarray}
    &&\rm{pair:}\ e^- + e^+ \longrightarrow \nu + \bar{\nu}, \\ 
    &&\rm{freep\ EC:}\ e^- + p \longrightarrow \nu_e + n, \label{nuc0} \\
    &&\rm{nuclei\ EC:}\ (Z,A) + e^- \longrightarrow (Z-1,A) + \nu_e, \label{nuc1}\\
    &&\rm{\beta^+:}\ (Z,A) \longrightarrow (Z-1,A) + e^+ + \nu_e, \label{nuc2}\\
    &&\rm{PC:}\ (Z,A) + e^+ \longrightarrow (Z+1,A) + \bar{\nu}_e, \label{nub3}\\
    &&\rm{\beta^-}:\ (Z,A) \longrightarrow (Z+1,A) + e^- + \bar{\nu}_e, \label{nuc4}
\end{eqnarray}
whereas other thermal processes such as plasmon decay, photo-neutrino processes and bremsstrahlung are neglected due to their small contributions for iron-CCSNe \citep{Kato2015,Guo2016,Patton2017b}.


The calculation procedures essentially follow those in \cite{Kato2017}. We calculate the reaction rate of pair annihilation by using the same formulation in  \cite{mezzacappa1993} and the same procedure in \cite{Kato2015}. The local spectra for $\nu_e$ and $\bar{\nu}_e$ are given by the integral of the reaction rate over the angle between two directions of the neutrino pair and the energy of the partner. For nuclear weak processes, on the other hand, we essentially take the same procedure in \cite{furusawa2017}, in which they employ literature tables FFN \citep{fuller1985}, ODA \citep{oda1994}, LMP \citep{langanke2001}, LMSH tables \citep{langanke2001c}, and adding the Tachibana (TB) table \citep{tachibana1995,yoshida2000,koura2004,koura2005}. Wherever tables overlap, we adopt the data from the table in the following preference: LMSH $>$ LMP $>$ ODA $>$ FFN for $\nu_e$ emission and LMP $>$ ODA $>$ FFN $>$ TB for $\bar{\nu}_e$ emission, respectively. We also employ an approximate formula provided by \citet{langanke2003} for the electron capture (EC) rate by heavy nuclei where reaction tables are unavailable. We also note that all reaction rates in the TB table are evaluated with Fermi-blocking of electrons taken into account by 
applying a suppression factor $1-f_e(\langle E_e \rangle)$ based on the average electron energy $\langle E_e \rangle$, which is given in the table. The mass fraction of nuclei, which are necessary quantities to evaluate weak reaction rates, is determined through two independent ways in this study: MESA fractions and NSE fractions. The former corresponds to the mass fraction obtained by nuclear network computations in stellar evolution simulations. The latter is, on the other hand, obtained by an assumption that nuclear statistical equilibrium  (NSE) is achieved: in practice we define the NSE region where $T>4\times10^{9}$ K and $\rho>10^{7.5} \rm{gcm^{-3}}$ are satisfied. We compute the NSE mass fraction with respect to $10^5$ nuclei \citep{furusawa2017b}. 
Our motivation for employing two different treatments on computing nuclear compositions are described in Section \ref{sec:NSE}.



The neutrino number spectrum and its energy-integrated quantity are computed by the volume integral of those of local quantities for all neutrino reactions. The integration is taken from the centre to some termination radius. In the case of MESA fractions, the integration is terminated at a radius where the increase rate becomes less than the total value by $10^{-8}$. 
In the case of NSE mass fractions, on the other hand, the integration is terminated at the outer edge of the NSE region.

\begin{figure}
    \centering
    \includegraphics[width=\columnwidth]{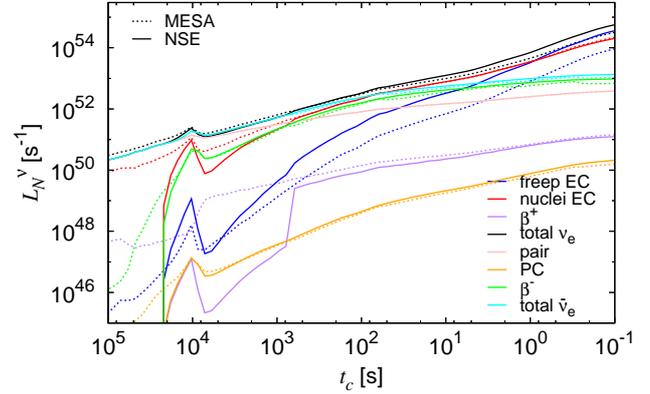}
    \caption{The time evolution of number luminosities for our base model, ($\Delta M_{\rm{max}},N_{\rm{net}}$) = (0.1, 204).
    Solid and dotted lines denote the results for NSE and MESA fractions, respectively. Different colors indicate different neutrino processes. The time $t_c$ 
    is defined as time to the onset of core collapse.
    }
    \label{dm0.1_net204}
\end{figure}

\begin{figure}
    \centering
    \includegraphics[width=\columnwidth]{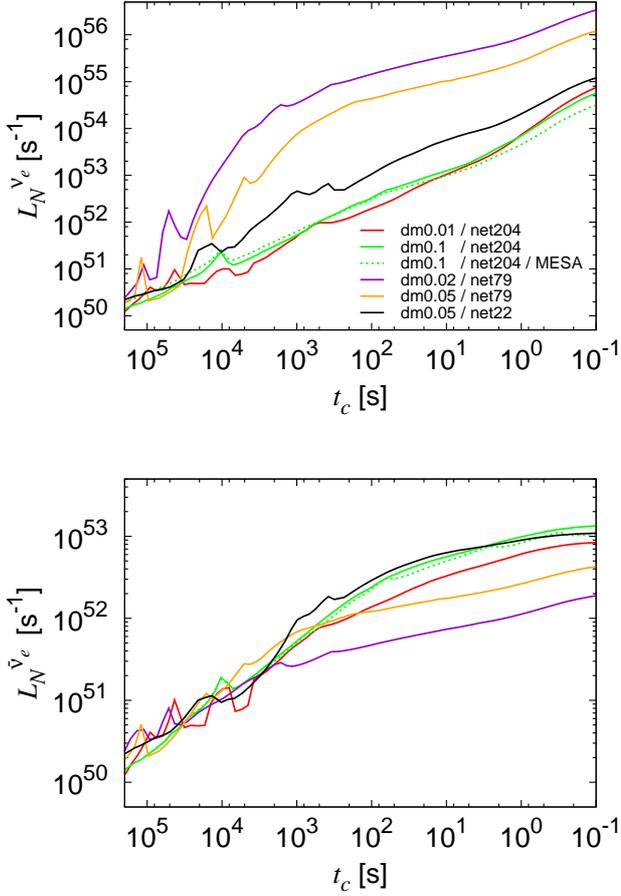}
    \caption{The time evolution of the neutrino number luminosities for the selected models with 15$M_\odot$.
    Top and bottom panels show the results for $\nu_e$ and $\bar{\nu}_e$, respectively.
    We basically employ NSE fractions in these calculations (solid), but we show results obtained with MESA fractions for the model with ($\Delta M_{\rm{max}},N_{\rm{net}}$) = (0.1, 204) for comparison (dotted).}
    \label{N_lum}
\end{figure}

\begin{figure}
     \centering
     \includegraphics[width=\columnwidth]{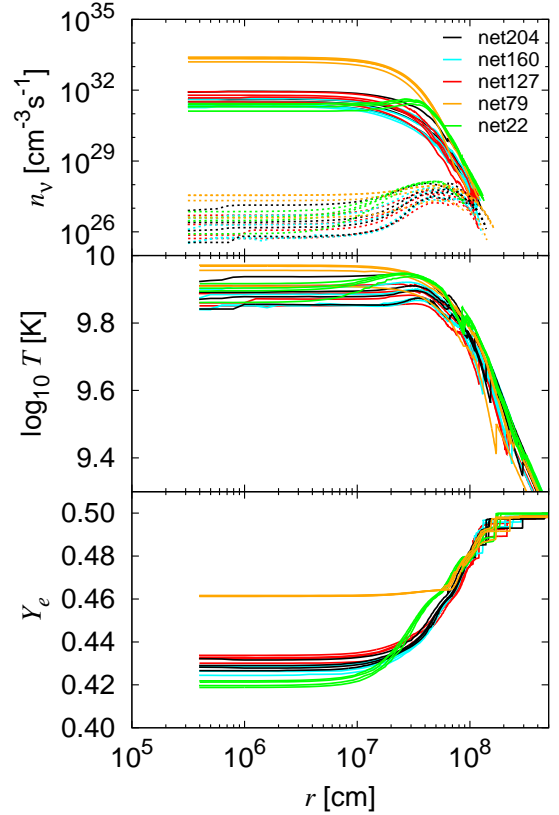}
     \caption{The radial profiles of neutrino emissivity (top), temperature (middle) and electron fraction (bottom) at the time when the central density has reached a threshold value ($\rho_c\sim10^{9.95}~\mathrm{g~cm^{-3}}$). We display them for all 20 progenitor models.   
     In the top panel, solid and dotted lines denote $\nu_e$ and $\bar{\nu}_e$, respectively.}
     \label{radial_prof}
 \end{figure}

\section{Results}
\subsection{Diversity of presupernova neutrinos} \label{sec:variety}


We first summarize some essential properties of presupernova neutrinos by showing the result of our base model, ($\Delta M_{\rm{max}},N_{\rm{net}}$) = (0.1, 204).
In Figure~\ref{dm0.1_net204} we show the time evolutions of number luminosities $L_N^\nu$ and the contributions from each weak-process. We postpone detailed discussion regarding the impact of different treatment of computations for nuclear compositions (MESA versus NSE fractions) to Section~\ref{sec:NSE}, and we focus on the result of the models with NSE fractions here.

For $\bar{\nu}_e$, neutrino emissions via pair annihilation and $\beta^-$ decay dominate other reactions, while EC reactions on nuclei and free protons make a significant contribution to the total number luminosity of $\nu_e$.
These are the same features as reported in previous studies \citep{Kato2017}.
The peaks associated with Si-shell burning appear in the number luminosities of both $\nu_e$ and $\bar{\nu}_e$ at $t_c \sim 10^4$ s. This is attributed to the fact that
the temperature inside the core decreases due to the core expansion with large energy generation through Si-shell burning and neutrino luminosities decrease accordingly \citep{Yoshida2016}.

Figure~\ref{N_lum} portrays the time evolutions of neutrino luminosities on different stellar evolution models, in which we pick up the results of 6 representative models: ($\Delta M_{\rm{max}},N_{\rm{net}}$) = (0.01, 204), (0.1, 204), (0.02, 79), (0.05, 79) and (0.05, 22).
The first one has the highest resolution and the largest number of nuclear species; thus, it may be the most realistic result.
However, we stress that the models have not reached numerical convergence and therefore the different spatial resolutions are only seeds for the stochasticity of the late stages of the evolution.
The others correspond to models computed with smaller nuclear network size than that of the first and second models. In this figure, we display the results of models calculated with NSE fractions. For comparison, the results for the base model calculated with MESA fractions 
are also included.




We find that the number luminosities differ among the models by more than an order of magnitude in both neutrino flavors, which reflects the diversity of stellar evolution models, indeed. As a general trend, models with higher temperature/electron fraction tend to have higher $\nu_e$ luminosities; for instance, the model with ($\Delta M_{\rm{max}},N_{\rm{net}}$) = (0.02,79) has the highest central temperature and electron fraction among models, which results in the highest $\nu_e$ luminosity.
As shown in the bottom panel of Figure~\ref{N_lum}, the $\bar{\nu}_e$ luminosity, on the other hand, follows the inverse trend to that of $\nu_e$; for example, the model with ($\Delta M_{\rm{max}},N_{\rm{net}}$) = (0.02,79), which corresponds to the model with the highest $\nu_e$ luminosity, has the weakest $\bar{\nu}_e$ emissions among models.

The response of $\nu_e$ and $\bar{\nu}_e$ to the difference of matter states can be understood through the property of the dominant weak process. For $\nu_e$ emissions, EC on heavy nuclei and free protons dominate other reactions (see Figure~\ref{dm0.1_net204}) and the increase of matter temperature and electron fraction facilitates these reactions. For $\bar{\nu}_e$ emissions, on the other hand, the dominant process is $\beta^-$ decay (see Figure~\ref{dm0.1_net204}) which tends to be suppressed with increasing electron fraction owing to the Fermi-blocking of electrons, whereas its temperature dependence remains positive. These two effects compete with each other in $\bar{\nu}_e$ luminosity for the model with ($\Delta M_{\rm{max}},N_{\rm{net}}$) = (0.02,79), though reduction by the effect of high electron fraction ends up winning.

Our results also suggest that presupernova neutrinos are more sensitive to the difference of the network size than that of spatial resolutions on stellar evolution simulations within the range explored. This is attributed to the fact that some important nuclear reactions are artificially suppressed by a shortage of some of the elements, which leads to qualitatively different matter profiles at the final stage of their evolution. Models with $N_{\rm{net}}$=22, for instance, have off-center temperature peaks (see the middle panel of Figure~\ref{radial_prof}), whereas its maximum locates at the center for other models. For models with $N_{\rm{net}}$=79, on the other hand, the central values of both temperature and electron fraction are remarkably deviated from those of other models, which plays a dominant role to enhance (reduce) the $\nu_e$ ($\bar{\nu}_e$) luminosity (see the top panel of Figure~\ref{radial_prof}). We note, however, that the matter distributions are qualitatively similar among models with $N_{\rm{net}} \geq 127$; accordingly the difference of neutrino luminosities is confined within the same order of magnitude. 
We conclude $N_{\rm{net}} \gtrsim 100$ is a must in stellar evolution simulations to capture the qualitative trend of presupernova neutrinos\footnote{This number depends on what the key reactions are for shaping the matter profile. It may only depend on a few key reactions but larger network sizes have higher possibilities to cover them.}.
\subsection{MESA versus NSE fractions} \label{sec:NSE}

\begin{figure}
    \includegraphics[width=\columnwidth]{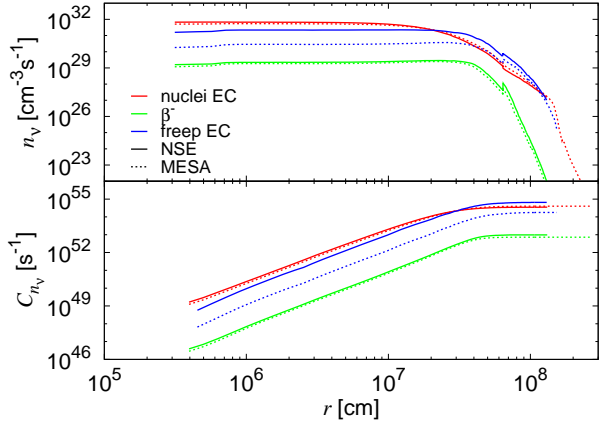}
     \caption{The radial profiles of neutrino emissivities (top) and the cumulative emissivities (bottom) for nuclei EC and $\beta^-$ decay at $t_c=0$~s of the base model.}
     \label{radial_neutrino}
\end{figure}

\begin{figure}
    \includegraphics[width=\columnwidth]{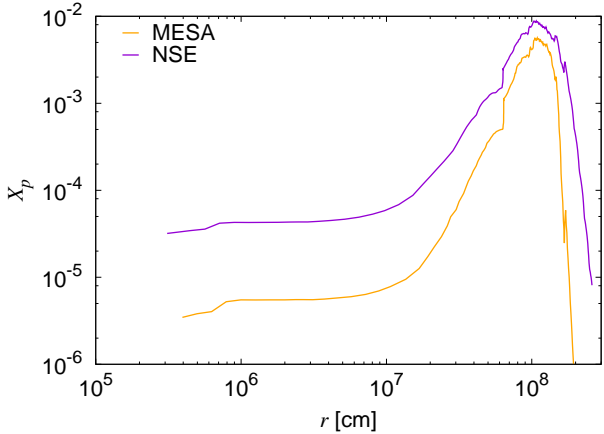}
     \caption{The radial profiles of proton mass fractions in the MESA (orange) and the NSE fraction model (purple) at $t_c=0$ s of the base model.}
     \label{pfrac}
\end{figure}

\begin{figure}
\includegraphics[width=\columnwidth]{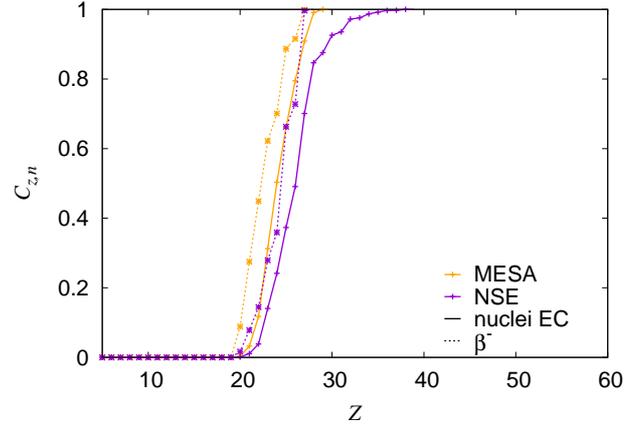}     \caption{The cumulative emissivities as a function of proton number $Z$ at the innermost point at $t_c=0$ s of the base model. \label{cum}}
\end{figure}

  \begin{figure*}
    \includegraphics[width=23cm,angle=-90]{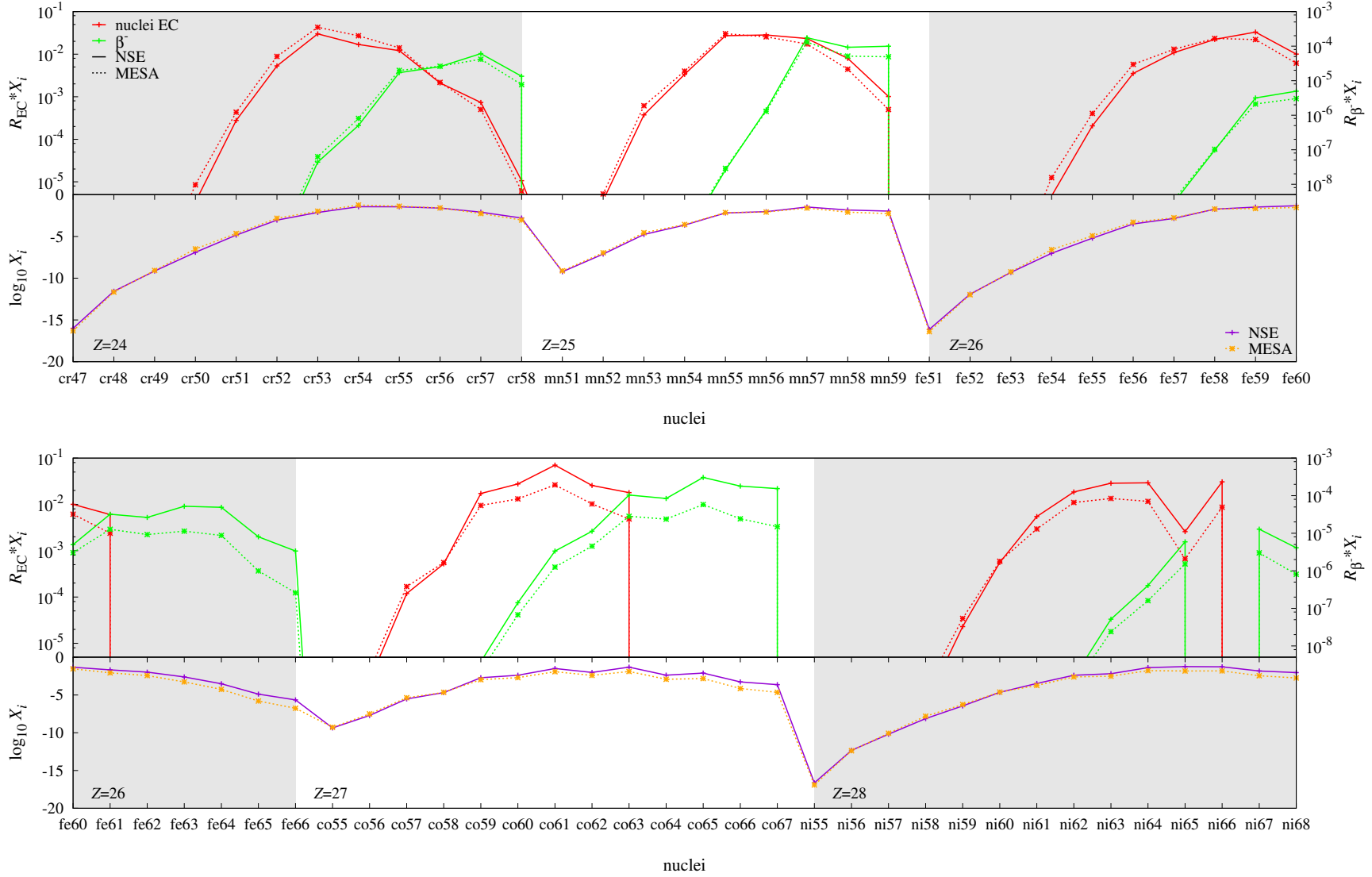}
     \caption{The products of mass fractions and reaction rates for 
     nuclei EC and $\beta^-$ decay (upper half) and mass fractions (lower half) for the nuclei, which make larger contributions to the neutrino emissivity.
     We take the results from the innermost part of the core at $t_c=0$ s of the base model.}
     \label{furusawa_comp}
 \end{figure*}

Before going into further details of the diversity of presupernova neutrinos, we quantify the impact of the different treatments in the computations of nuclear composition in this section. For readers that are not familiar with the literature, we first explain the reason why this study is important.

As we have already mentioned, NSE, in which the forward and reverse reactions on the nuclear network are
balanced, is fulfilled around the center at the last stage of stellar evolution, and the weak processes in the NSE region dominates the neutrino emission at this phase (see below for more details). Given the NSE condition, we can obtain the nuclear composition without solving nuclear reaction networks, i.e., its computation is less expensive than that of solving full reaction networks. This allows us to set a large number of nuclei in the computation of nuclear composition; indeed, our NSE computations takes into account $10^{5}$ nuclei, which includes the full nuclear composition associated with the stellar evolution. In the case with MESA fractions, on the other hand, the composition is determined through computations of nuclear reaction networks, where the computation depends strongly on the network size and it becomes very expensive for $N_{\rm{net}} > 200$. It is confirmed, however, that all nuclear reactions around the center settle into a detailed balanced state, implying that NSE is also achieved in the MESA model. This may sound as though our NSE and MESA fractions may look alike, which is not true, though. The crucial problem is the small network size in MESA simulations where the maximum number is 204. This is much smaller than that of our NSE model and such a limited number of network size leads to a different equilibrium state, i.e., the obtained nuclear composition may deviate from that in real NSE \footnote{Note that nuclear composition in NSE depends on the details of the nuclear model such as shell effects \citep[see e.g.][]{furusawa2017}. As such, the uncertainties of nuclear models are also inherited to this study.}. As we shall see below, the deviations between MESA and NSE fractions are actually observed even in the largest network size in our MESA model. This issue is not directly connected to the uncertainty of stellar evolution models but rather being an independent uncertainty to cause the diversity of presupernova neutrinos; hence we quantify the impact in this section.



We find that the difference of neutrino luminosities in the cases between our NSE and MESA fractions on the base model ($\Delta M_{\rm{max}},N_{\rm{net}}$) = (0.1, 204) is $\sim$ 35\% and $\sim$ 15\% for $\nu_e$ and $\bar{\nu}_e$, respectively (see also green solid and dotted lines in Figure~\ref{N_lum}). For $\nu_e$, the difference mainly comes from the difference in EC on free protons, which is a factor $\sim 10$, whereas the differences of EC on heavy nuclei are less remarkable ($\sim 10 \%$) (see the top panel of Figure~\ref{radial_neutrino}). It is also interesting to point out that the most dominant weak process for $\nu_e$ emission is different between the MESA and NSE models. To see this more quantitatively, we compute (radially) cumulative emissivity, defined as
\begin{eqnarray}
C_{n_\nu}(r) = \int_0^{r} 4\pi r^{\prime2} n_\nu(r^\prime) dr^\prime,
\end{eqnarray}
and the results are shown in the bottom panel of Figure~\ref{radial_neutrino}. For the case with NSE fractions the most dominant weak process is EC on free protons, whereas it is EC on heavy nuclei for the case with MESA fractions. This result suggests that the appropriate treatment of NSE is mandatory; otherwise even the dominant process can be obscured.

One of the interesting facts in the above analysis is that EC on free protons is more sensitive to the different treatments of nuclear composition than that of EC on heavy nuclei. This is attributed to the fact that the mass fraction of free protons ($X_p$) depends strongly on the treatment of nuclear composition; indeed the difference of radial distribution of $X_p$ between NSE and MESA is remarkable (see in Figure~\ref{pfrac}). Note that the neutrino emissions of EC on free protons can dominate other reactions even if the mass fraction is less than $10^{-2}$, since its emissivity per nucleon is the highest among the weak processes.

It is somewhat complicated to understand the quantitative difference of EC on heavy nuclei between MESA and NSE models, since EC on many nuclei are involved; thus we first attempt to narrow down the range of nuclei which contributes dominantly. To do so, we compute the cumulative emissivities at the stellar center as a function of proton number $Z$, which is defined as 
\begin{eqnarray}
C_{Z,n} = \frac{1}{n_\nu}\sum_{i=1}^Z n_{\nu,i}\ ,
\end{eqnarray}
where $n_{\nu,i}$ denotes the neutrino emissivity of EC on heavy nuclei with a proton number $i$, summed over all the isotopes. Note that $C_{Z,n}$ is normalized by the total $n_\nu$, i.e., its maximum is 1.
The result is shown in Figure~\ref{cum} (see solid lines). We find that $C_{Z,n}$ rises up at $Z \sim 20$ and then reaches almost unity at $Z \sim 30$ for both cases, implying that the nuclei with $Z$ = 20 -- 30 dominate the contributions of EC on heavy nuclei. We also find a different trend between NSE and MESA fraction models; $C_{Z,n}$ on MESA model increases sharper than that on NSE model, which indicates that the mass fractions of lighter nuclei in MESA model tends to be higher than that of NSE model. To see the trend more quantitatively, we summarize the distribution of $X_i$ and $X_iR_i$ for each nucleus in Figure~\ref{furusawa_comp}. 
Here the $X_i$ and $R_i$ denote the mass fractions of nuclei and the reaction rates of EC and $\beta^-$ decay, respectively.
As shown in the plots, the mass fraction of lighter nuclei is slightly larger in MESA than the NSE model. The excess of light nuclei in the MESA model may also be associated with the small mass fraction of protons. 
On the other hand, the neutrino emissivities at the heavier nuclei in the NSE models are higher than those in MESA. The two effects compete each other in total neutrino emissions, and eventually the latter effect wins, i.e.,  the neutrino luminosity becomes higher in the model with NSE fraction than that of MESA one (see the upper-half panels).

We now turn our attention to $\bar{\nu}_e$, which has $\sim 15 \%$ difference between the NSE and MESA cases. Although the difference is smaller than that of $\nu_e$, we briefly look into the origin. The dominant weak process is $\beta^-$ decay of heavy nuclei, hence we perform a similar analysis as the one for EC on heavy nuclei. The qualitative trends of the cumulative neutrino emissivity is very similar as that in EC of heavy nuclei (see Figure~\ref{cum}): the heavy nuclei with $Z$ = 20 -- 30 plays a dominant role in neutrino emissions by $\beta^-$ decay: the MESA model has a sharp increase of  $C_{Z,n}$ because of the excess of lighter nuclei. The detailed information on neutrino emissivities by $\beta^-$ decay of each nucleus is shown in Figure~\ref{furusawa_comp}. We find that the mass fractions for the neutron rich isotopes of Fe, Co and Ni are different by a factor $\lesssim$13 between the two models, which plays the most dominant role to account for the difference of $\bar{\nu}_e$ luminosity between NSE and MESA cases.

Finally we make a comment on the contribution of neutrino emissions from non-NSE regions. As shown in the bottom panel of Figure~\ref{radial_neutrino}, the cumulative emissivity is almost constant at $r \gtrsim 3 \times 10^{7}$ cm for all weak processes, indicating that neutrino emissions in the non-NSE region has a negligible contribution. However, we warn that this is not true in the earlier phase. For instance, neutrino emissions in non-NSE region dominates those of NSE region during the Si-shell burning phase ($t_c \sim 10^{4}$ s). It should be noted, however, that the neutrino emissions are dominated by thermal processes such as electron-positron pair annihilation, hence the effects by the different treatments of nuclear composition are subtle at this phase.


\section{Phenomenological model} \label{sec:correlation}

\begin{figure*}
    \centering
    \includegraphics[width=14cm]{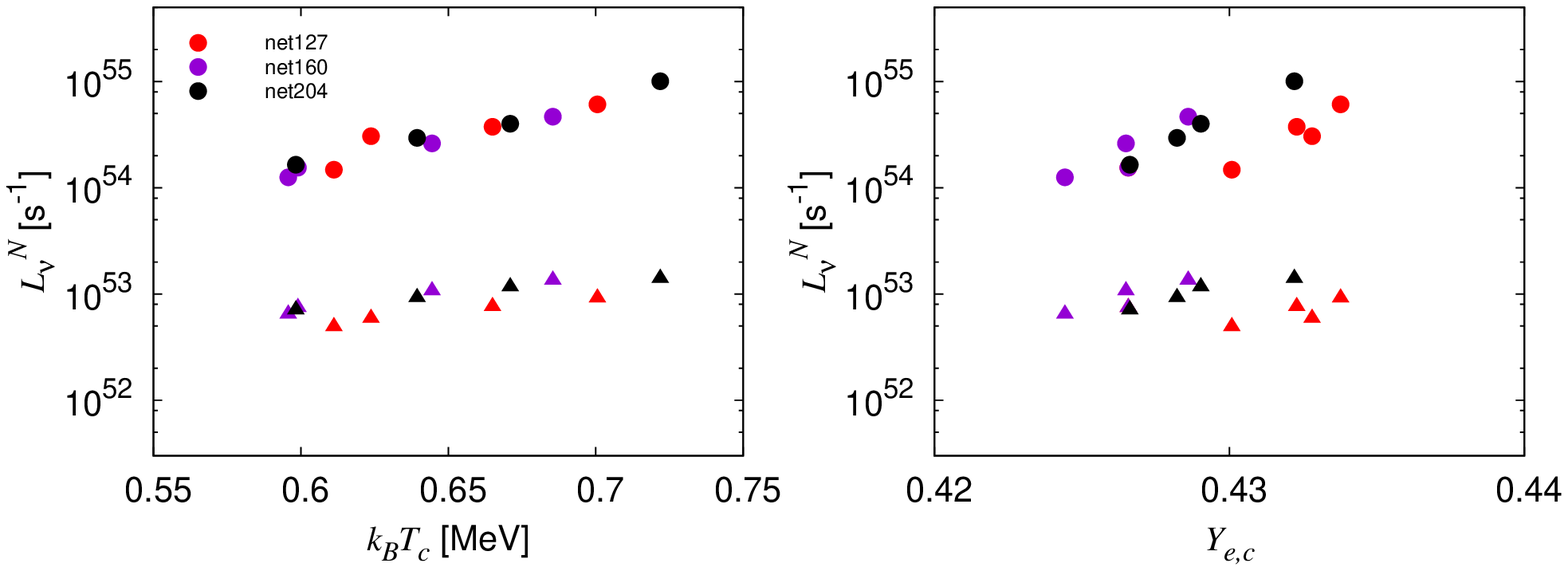}
     \includegraphics[width=19cm]{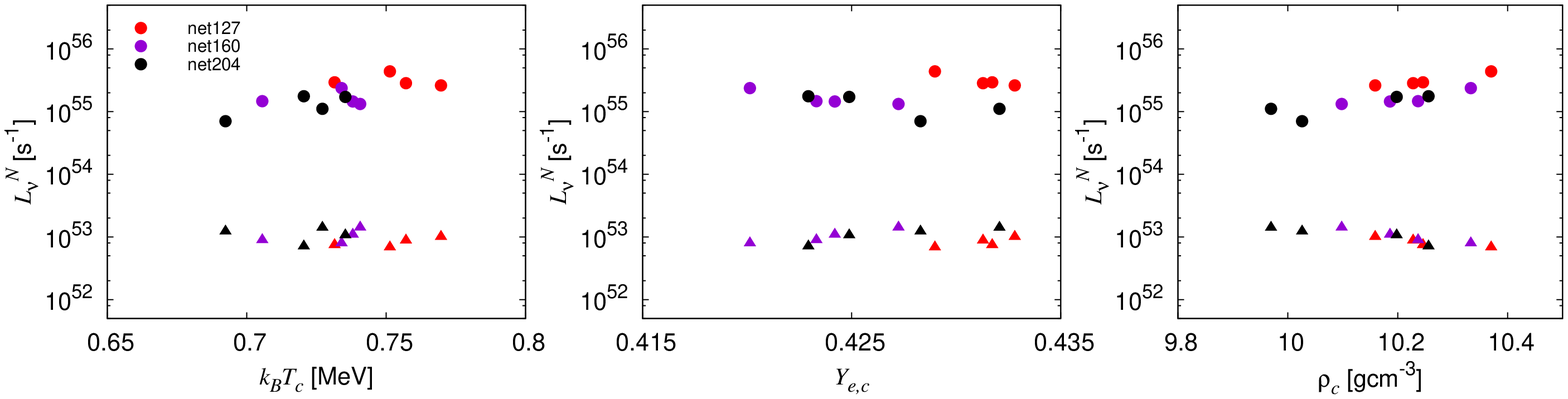}
    \caption{Scatter diagrams of number luminosities with the central quantities for the progenitor models with $N_{\rm{net}}\geq127$ at two different time snapshots, $\rho_c \sim 10^{9.95}\ \rm{gcm^{-3}}$ (top) and $t_c=0$ s (bottom). Circles and triangles denote $\nu_e$ and $\bar{\nu}_e$, respectively. Different colors indicate the differences of the network size employed in the stellar evolution models.}
    \label{correlation}
 \end{figure*}

As described in Section~\ref{sec:variety}, we reveal a rich diversity of presupernova neutrinos induced by different stellar evolution models and it can be understood through the response of the dominant weak process to the differences in matter state. These analyses are meaningful to clarify the origin of the diversity, whereas they may not be useful for real observations. In reality, spatially integrated neutrino emissions will be the only observable quantity, which implies that the information on the spatial distribution is limited. Our previous detailed analyses on the response of neutrino luminosities through weak processes focused on local processes and the same methodology cannot be applied directly to global quantities. To understand the global trend, we need to approach the problem in a different way, for example, by developing phenomenological models.
In this section, we perform a novel analysis to address this issue with paying particular attention to relations between the central matter quantity and neutrino luminosities.




We start out by summarizing the relation between neutrino luminosities and three central matter quantities: density, temperature and electron fraction,
which will be the only independent variables that determine the neutrino luminosity in our phenomenological model.
The results are shown in Figure~\ref{correlation}, where we select two different time snapshots: the time when the central density reaches a threshold value $\rho_c \sim 10^{9.95}\ \rm{gcm^{-3}}$ (top panels) and $t_c=0$ s (bottom panels).
Note that we only focus on models with $N_{\rm{net}}\geq127$ here, since others seem to be unrealistic (see Section~\ref{sec:variety} for more details).
Although some clear correlation can be seen (for instance, between $\nu_e$ luminosities and the central temperature), they are less prominent among almost every quantity. 
It should be also emphasized that $\bar{\nu}_e$ seems to have a negative correlation to the central electron fraction (see triangles in the top-right and bottom-middle panels of Figure~\ref{correlation}), which is an unexpected result from the argument of response of $\beta^-$ decay (this the most dominant emission process for $\bar{\nu}_e$) to electron fraction (see Section~\ref{sec:variety}).
At first glance, those central quantities are not adequate to characterize presuppernova neutrinos, which turns out to be not true, through.

A crucial issue in the above argument is that the effects of other quantities are also involved in the comparison, which smears out the correlation. For instance, when we look into the correlation between neutrino luminosities and central temperature at fixed density $\rho_c \sim 10^{9.95}\ \rm{gcm^{-3}}$ (e.g., the top-left panel of Figure~\ref{correlation}), the central electron-fractions are also different among models. As a consequence, the neutrino luminosities do not reflect only the difference of temperature but also that of electron fraction as well. Although we are able to narrow down the independent variables from three to two by simply changing the time snapshot, it is impossible to further simplify the problem into a single-variable problem in the same way.

Our phenomenological model provides an approximate solution to overcome the above issue. The guiding principle of our approach is that neutrino luminosities can be approximately expressed as a function of central density, temperature and electron fraction, i.e., $L_\nu^N \sim L_\nu^N(\rho_c,T_c,Y_{e,c})$. We accept the fact that this approach discards some detailed information on spatial distributions of neutrino emissions.
However, the simplification makes it possible to compare neutrino luminosities to central values without being obscured by other effects, which is demonstrated below.



The detail of the method is as follows. We assume that $L_\nu^N$ is a smooth function of central density, temperature and electron fraction; hence its logarithm, $\log_{10}{L_\nu^N}$, can be expanded as a Taylor series to linear order as
\begin{eqnarray}
 &&\log_{10}{L_\nu^N} \left( \rho, T, Y_{e} \right) \sim  \log_{10}{L_\nu^N} \left(\rho_{0}, T_{0}, Y_{e0}\right) \nonumber \\
 &&+  \left. \frac{\partial \log_{10}{L_\nu^N}}{\partial \rho} \right|_{\substack{T=T_0\\Y_e=Y_{e0}}} \left(\rho - \rho_{0} \right)
 + \left. \frac{\partial \log_{10}{L_\nu^N}}{\partial T} \right|_{\substack{\rho=\rho_{0}\\Y_e=Y_{e0}}} \left(T - T_{0} \right) \nonumber \\
 && + \left. \frac{\partial \log_{10}{L_\nu^N}}{\partial Y_{e}} \right|_{\substack{\rho=\rho_{0}\\T=T_{0}}} \left(Y_{e} - Y_{e0} \right).
 \label{LT}
\end{eqnarray}
In the expression, we omit the subscript ``$c$'' with respect to each matter quantity, meanwhile the subscript ``0'' denotes the point that obtained from each stellar evolution simulation. By virtue of the expression, we can correct $L_\nu^N\left(\rho_{0}, T_{0}, Y_{e0}\right)$ to that at arbitrary different central quantities by giving derivative coefficients in Eq.~(\ref{LT}).
The derivatives can be evaluated by the following manner. Given the matter distributions computed by stellar evolution simulations, we artificially change each matter quantity by $\pm1\%$ for the entire spatial region, and then compute neutrino luminosity with the same procedure as described in Section~\ref{sec:neutrino} (the nuclear composition is assumed to be NSE). Using the obtained neutrino luminosities, we compute the derivative by the finite difference.

Below we apply the phenomenological analysis to our 12 models with $N_{\rm{net}} \geq 127$ at the time when $\rho_c \sim 10^{9.95}\ \rm{gcm^{-3}}$. Here we give two important remarks. The first one is that we do not include the models with $N_{\rm{net}} =$ 22 and 79, since their progenitor structures are qualitatively different from those with larger network size and they are unrealistic (see Section~\ref{sec:variety} for more details). The second one is that we apply the phenomenological analysis to neutrino luminosities computed at the time when the central density is the same among all models, implying that the density derivative term in Eq.~(\ref{LT}) is not necessary to correct $L_\nu^N$. 
To clearly see the correlations between neutrino luminosities and each variable,
we remove the contribution of one of the variables by shifting the value of the other variable to a reference value using Eq.~(\ref{LT}). For example, to see the neutrino luminosity as a function of central temperature, we shift the central electron fraction values to a reference value (our reference model is the model with $(\Delta M_{\rm{max}},N_{\rm{net}})=(0.01,204)$) and change the luminosity accordingly following Eq.~(\ref{LT}). We also take the same procedure vice versa.

\begin{figure*}
    \includegraphics[width=18cm]{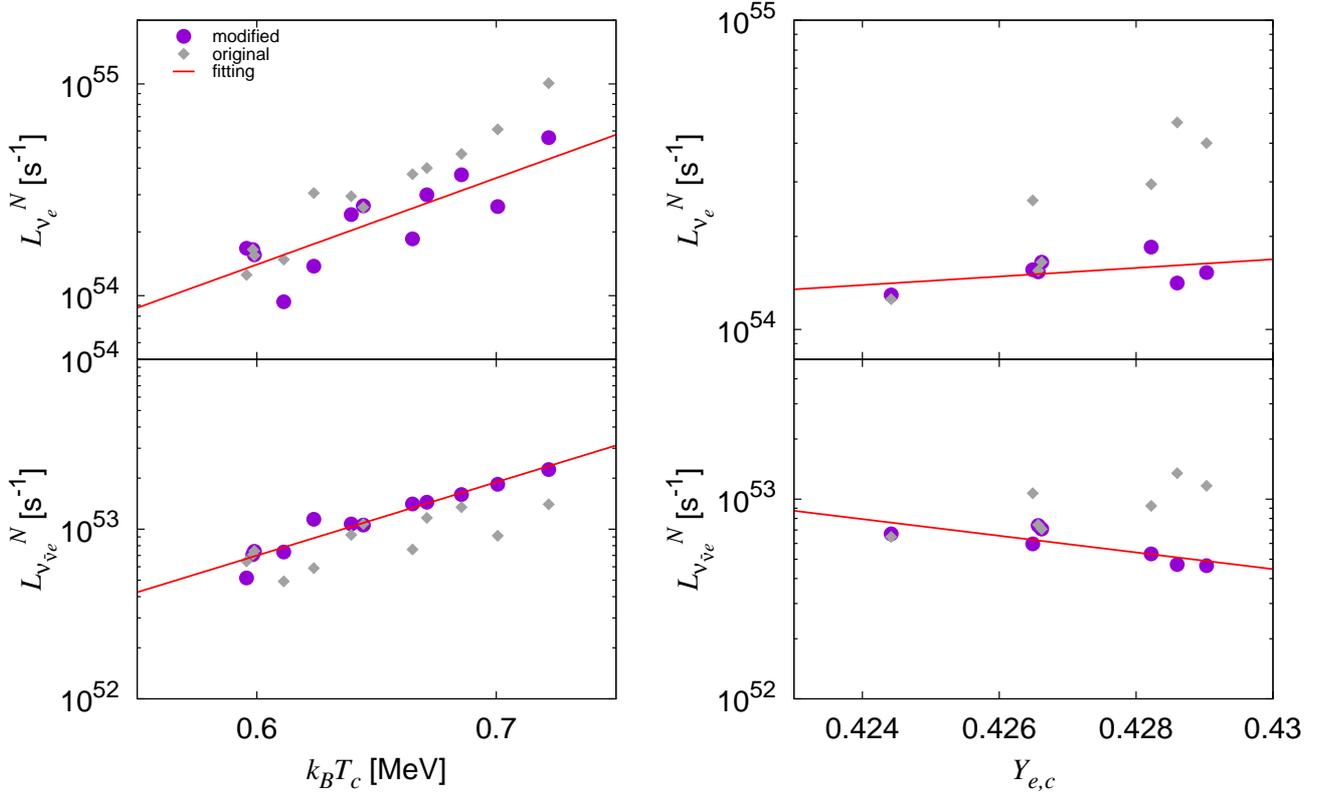}
     \caption{Scatter diagrams of the corrected number luminosities following Eq.~(\ref{LT}) (see the text for more details). The relations between temperature and neutrino luminosities are shown in the left panels, while those of electron fraction are shown in the right panels. The top and bottom panels represent the results for $\nu_e$ and $\bar{\nu}_e$, respectively. For comparison, we also show the original results (not corrected by  Eq.~(\ref{LT})) as grey diamonds. The red line is a linear fit to the corrected number luminosities.}
     \label{corelation_rhofix}
 \end{figure*}

The results are summarized in Figure~\ref{corelation_rhofix}. As shown in the figure, some remarkable correlations emerge, albeit somewhat scattered: both $\nu_e$ and $\bar{\nu}_e$ have a positive correlation to temperature, whereas the correlation to the electron fraction depends on the species; it is positive and negative for $\nu_e$ and $\bar{\nu}_e$, respectively. These trends are consistent with those expected from the arguments with weak processes (see Section~\ref{sec:variety} for more details); hence it 
seems to be real. 
This result suggests that our phenomenological model appropriately captures the expected response of neutrino luminosities to the differences of matter state. We also emphasize that the response reflects the global characteristics (not local one) of progenitors. 
 
 
 There is another advantage in the phenomenological approach, which is the most important ingredient in the analysis. Each derivative term in Eq.~(\ref{LT}) reflects a degree of response to the change of each matter quantity, indicating that we can quantify the influence of each matter quantity to neutrino luminosities. It should be emphasized that this is an important virtue of the phenomenological model; in fact, it is very hard to see what matter quantity is the most influential to neutrino luminosities just by using the raw results of our simulations. In the analysis without
 the phenomenological model, we need to carefully look into the radial profile of neutrino emissions and then determine what spatial region is the most dominant for them. The region would, however, depend on the model and, more importantly, we need to analyze the response of each weak processes with appropriate averaging prescriptions over the region, which is very complicated in practice. In our approach, on the other hand, we treat the central matter quantities as representative variables to characterize neutrino luminosities
 and allows us to measure the influence from each of them to neutrinos without any complicated procedure.

\begin{figure*}
    \includegraphics[width=18cm]{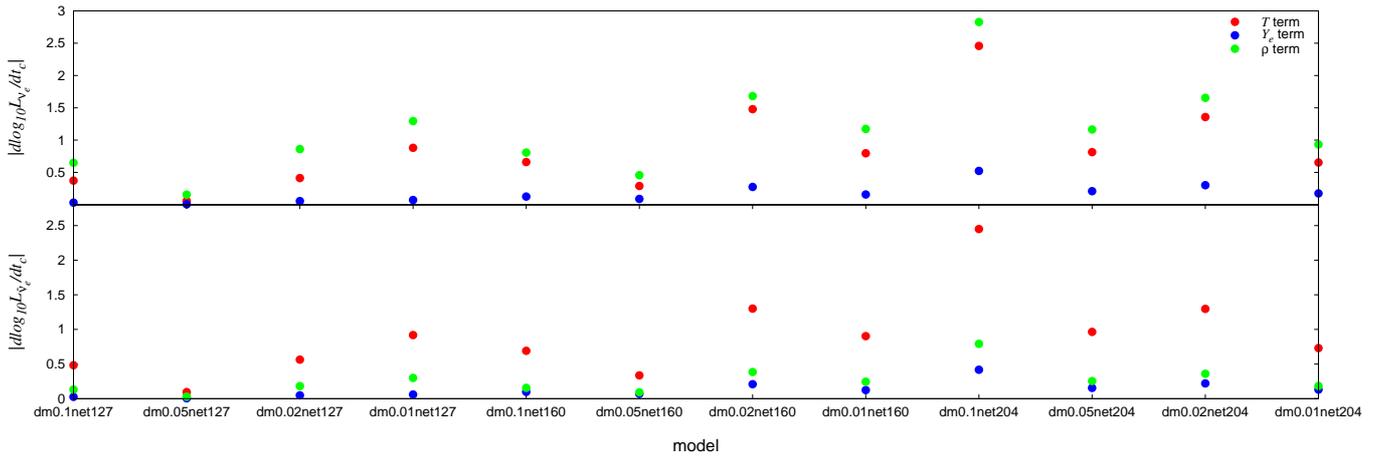}
     \caption{The absolute time derivatives of number luminosities with respect to the central density, electron fraction and temperature.
     The vertical quantities are defined by the product of each term of Eq.~(\ref{LT}) and the time derivative of the central quantities.
     The upper and lower panels show the results for $\nu_e$ and $\bar{\nu}_e$, respectively.
     We measure the time $t_c$ from the onset core collapse.}
     \label{correlation_dt}
 \end{figure*}

The results are summarized in Figure~\ref{correlation_dt}. We measure the sensitivity of neutrino luminosity to each central matter quantity in the same unit by multiplying the time derivatives of each of them and then taking their absolute values. For instance, the $T$ term is described as $|d\log_{10}L_\nu/dT\times dT/dt$|.
Those time derivatives are computed by using neutrino luminosities at the most nearby time snapshot on each stellar evolution model. As shown in the figure, we find that the density and temperature is the most influential quantity for $\nu_e$ and  $\bar{\nu}_e$, respectively. This implies that, for example, if we are able to observe the light curve of $\bar{\nu}_e$, the changes will be governed by the change in central temperature. The insensitivity of neutrino luminosities to the electron fraction is attributed to the fact that the electron fraction evolves slowly compared to density and temperature. It should be also emphasized that all of our results presented in the figure has a common trend, implying that the result is robust. It is very interesting to see whether the trend is
common for different mass progenitors, which will be addressed in our forthcoming paper, though.


 \section{Summary and discussion} \label{summary}
 In this paper, we investigate the diversity of presupernova neutrinos induced by different stellar evolution models for a 15$M_\odot$ progenitor.
 Among our models, the evolution starts to separate after Si-burning, and then the central density, temperature and electron fraction at the onset of core collapse vary in the ranges of $10^{9.96} \leq \rho_c/\rm{gcm^{-3}} \leq 10^{10.37}$, $0.674 \leq T_c/\rm{MeV} \leq 0.861$ and $0.418 \leq Y_e \leq 0.461$, respectively.
 This makes the number luminosities of presupernova neutrinos diverse.
 We calculate the time evolution of the neutrino number luminosities and find that the luminosites differ by more than an order of magnitude in both $\nu_e$ and $\bar{\nu}_e$.
 For $\nu_e$, the higher temperature/electron fraction models have the higher number luminosities, whereas it seems to be a reverse trend for $\bar{\nu}_e$. 
 The dependence of neutrino luminosities on each hydrodynamical quantity can be understood through the characteristics of the dominant weak processes as EC reactions and $\beta^-$ decay for $\nu_e$ and $\bar{\nu}_e$, respectively. We also find that the neutrino luminosities in models with smaller network size ($N_{\rm{net}}$ = 22, 79) are qualitatively different from those of higher network size. This is attributed to the fact that stellar evolution simulations with insufficient network size is not eligible to capture the nuclear burning accurately, which results in qualitatively different matter profiles at the final stages of the evolution. We hence conclude that $\gtrsim 100$ network size is mandatory in computations of stellar evolution to evaluate presupernova neutrinos accurately.
 
 We also investigate how uncertainties of nuclear compositions affect the diversity of presupernova neutrinos by comparing models between MESA and NSE fractions.
 We find that the number luminosities for both models differ by $\sim35$\% and $\sim15$\% for $\nu_e$ and $\bar{\nu}_e$, respectively, for the base model. For $\nu_e$, this is mainly
 due to the large deviation in EC on free protons where the proton mass fractions in the NSE model is $\sim 8$ times larger than that in the MESA model.
 This result indicates that the mass fraction of free protons is the most sensitive to the difference of computation methods for nuclear compositions, which has a strong impact on neutrino luminosities. Indeed, we find that the dominant weak process for $\nu_e$ is different between NSE and MESA models.
 The differences in the $\nu_e$ emissions by EC on heavy nuclei and $\bar{\nu}_e$ by $\beta^-$ decay are not so large, on the other hand, which are $\sim12$\% and $\sim25$\% 
 , respectively. We also find that the neutrino emissions in the non-NSE region can be negligible at the onset of collapse, although they can be dominant at earlier phases.
 
 We develop a phenomenological model to capture essential trends in the diversity of presupernova neutrinos aiming to be applied in future CCSN neutrino observations. The model is built based on the assumption that central quantities such as density, temperature and electron fraction can approximately characterize the property of neutrino luminosities, which in general should be dependent on the detailed stellar structure. We also quantify the influence of neutrino luminosities from each hydrodynamical variable by using our phenomenological model, and find that the density and temperature are the most influential quantity for $\nu_e$ and $\bar{\nu}_e$, respectively. Such a qualitative trend may be useful to interpret the future neutrino observations, which indicates that the phenomenological analysis will play a pivotal role to bridge the gap between the theory of stellar evolution and neutrino observations.

Finally we make some important remarks. The results presented in this paper are still incomplete and this paper is a pilot study for more thorough investigations to connect the presupernova neutrinos to the theory of stellar evolution. The top priority is to extend this work to different mass CCSN progenitors. In fact, some previous studies have revealed that the property of presupernova neutrinos depend on the mass of progenitor \citep[see e.g.][]{Kato2017}; hence, it should be addressed whether our conclusions hold for different masses. Another important aspect regarding the extension of this work is to apply the same analysis for a slightly earlier phase than that of this study. In this study, we mainly focus on the phase right before the onset of collapse, meanwhile presupernova neutrinos may be observed from the Si-shell burning phase. Such a study is definitely required to develop a comprehensive understanding of the connection between presupernova neutrinos and the final stages of stellar evolution, all of which will be presented in our forthcoming papers.

 \section*{Acknowledgements}
 We are gratefully thanks to Dr. Furusawa for providing the calculation code of NSE mass fractions.
 C.K. is supported by Tohoku University Center for Gender
Equality Promotion (TUMUG) Support Project (Project to Promote Gender Equality and Female Researchers).

\bibliographystyle{mnras}
\bibliography{ref.bib}







\bsp	
\label{lastpage}
\end{document}